\documentclass[fleqn,usenatbib,useAMS]{mnras}

\usepackage{graphicx}	
\usepackage{amsmath}	
\usepackage{amssymb}	
\usepackage{multicol}        
\usepackage{bm}		
\usepackage{pdflscape}	
\usepackage{array}
\usepackage{fancyvrb}
\usepackage{subfig}
\usepackage{enumitem}

\usepackage[T1]{fontenc}
\usepackage{ae,aecompl}
\usepackage{txfonts}
\usepackage{threeparttable}
\usepackage{breakurl}
\usepackage{longtable}
\usepackage{adjustbox}
\usepackage{longtable,csvsimple,booktabs}
\usepackage{xcolor}

\title[A Machine Learning Approach For Classifying Low-mass X-ray Binaries]{A Machine Learning Approach For Classifying Low-mass X-ray Binaries Based On Their Compact Object Nature}

\author[R. Pattnaik et al.]{
R. Pattnaik,$^{1,2}$\thanks{E-mail: rp2503@rit.edu}
K. Sharma,$^{3,4}$
K. Alabarta,$^{2,5}$
D. Altamirano,$^{2}$
M. Chakraborty$,^{6}$ \newauthor
A. Kembhavi,$^{4}$
M. Mendez$^{5}$
and J.K. Orwat-Kapola$^{2}$
\\
$^{1}$School of Physics and Astronomy, Rochester Institute of Technology, Rochester, NY 14623, USA\\
$^{2}$School of Physics and Astronomy, University of Southampton, Southampton, Hampshire SO17 1BJ, UK\\
$^{3}$Aryabhatta Research Institute of Observational Sciences (ARIES), Manora Peak, Nainital 263001, India\\
$^{4}$Inter University Centre for Astronomy and Astrophysics (IUCAA), Pune 411007, India\\
$^{5}$Kapteyn Astronomical Institute, University of Groningen, P.O. BOX 800, 9700 AV Groningen, The Netherlands\\
$^{6}$DAASE, Indian Institute of Technology Indore, Khandwa Road, Simrol, Indore-452020, M.P., India\\
}

\date{Accepted 2020 December 11. Received 2020 December 04; in original form 2020 April 16}

\pubyear{2020}
\hypersetup{draft}

\begin{document}

\label{firstpage}
\pagerange{\pageref{firstpage}--\pageref{lastpage}}
\maketitle

\begin{abstract}

Low Mass X-ray binaries (LMXBs) are binary systems where one of the components is either a black hole or a neutron star and the other is a less massive star. It is challenging to unambiguously determine whether a LMXB hosts a black hole or a neutron star. In the last few decades, multiple observational works have tried, with different levels of success, to address this problem. In this paper, we explore the use of machine learning to tackle this observational challenge. We train a random forest classifier to identify the type of compact object using the energy spectrum in the energy range 5-25 keV obtained from the Rossi X-ray Timing Explorer archive. We report an average accuracy of 87$\pm$13\% in classifying the spectra of LMXB sources. We further use the trained model for predicting the classes for LMXB systems with unknown or ambiguous classification. With the ever-increasing volume of astronomical data in the X-ray domain from present and upcoming missions (e.g., SWIFT, XMM-Newton, XARM, ATHENA, NICER), such methods can be extremely useful for faster and robust classification of X-ray sources and can also be deployed as part of the data reduction pipeline.

\end{abstract}




\section{Introduction}\label{sec:intro}

Low Mass X-ray binaries (LMXBs) are binary systems where one of the components is a black hole (BH) or a neutron star (NS) and the other component is a less massive star, usually a main sequence, a white dwarf or an evolved star of $M<1M_{\odot}$. Some LMXBs combine long periods of quiescence (from a few months to decades) with short periods where the source is in outburst that last from days to years. In quiescence, LMXBs are very faint ($\sim10^{30}$-$10^{33}$ erg s$^{-1}$), while during outbursts LMXBs increase several orders of magnitude their fluxes \citep[see, e.g.][]{McClintock2006}. 

The energy spectra of LMXB systems are described by two main components: a thermal component and a hard component. The thermal component is usually described by a multi-colour disc blackbody \citep{Mitsuda1984} and it is thought to be produced by an accretion disc \citep{Shakura1973}. The hard component is thought to be produced by the so-called corona, which is a region of hot plasma around the compact object \citep[e.g., ][]{Sunyaev1980}. This component is usually described by a thermal Comptonisation model \citep[e.g., ][]{Titarchuk1994, Done2007}. The contribution of these components to the X-ray emission of LMXBs varies during an outburst, modifying its spectral and timing properties \citep[e.g., ][]{VanDerKlis1989, Mendez1997, Homan2005, Remillard2006, Belloni2010, tetarenko2016watchdog}. LMXB show different spectral states during an outburst based on its spectral and timing properties \citep[e.g., ][]{Homan2005, Remillard2006, Belloni2010}. The two main states are the high/soft state (hereafter HSS) and low/hard state (hereby LHS). In the HSS the accretion disc is thought to extend down to the surface of the NS or the last stable orbit (if the compact object is a BH). Because of that the energy spectrum is dominated by the accretion disc, which is described by the thermal component. In the LHS the disc is thought to be truncated at larger radius than in HSS, so the spectrum is dominated by the corona, usually described by the Comptonised component. Between these two spectral states, the source can show different intermediate states with spectral and timing properties between the properties of the LHS and HSS. The evolution along these states can be well studied with the hardness-intensity diagram \citep[HID, see, e.g., ][]{Homan2001} and the colour-colour diagram \citep[CCD, eg., ][]{Hasinger1989, VanDerKlis1989}.

One of the fundamental questions when studying (LMXBs) is whether the compact object in the binary is a NS or a BH. The presence of one or the other can have a significant impact in the physical interpretation of the phenomenology observed. With the large scale sky surveys 
and transient search programs (e.g.,  \textit{INTEGRAL}/JEM-X \citep{Lund2003}, \textit{Swift}/BAT Transient Monitor \citep{Krimm2013}, \textit{MAXI} \citep{Matsuoka2009}, eROSITA \citep{Merloni2012}), the sample of LMXBs is ever-increasing. Such newly detected transient sources are usually characterised by their fast variation (days) of luminosity by orders of magnitude. The early identification of the nature of the compact object is very important for the community to be able to trigger expensive (and usually difficult to plan) observing campaigns \citep[][]{Middleton2016}. These campaigns, in most cases, can only be triggered if the nature of the compact object is known. There are only a few methods that allow the community to unambiguously identify the nature of the compact object: coherent pulsations \citep[][and references therein]{Patruno2012} and presence of thermonuclear bursts \citep[for reviews, e.g., ][]{Lewin1993, Cumming2004, Galloway2008, Strohmayer2018}, which determines unambiguously that the compact object is a NS, and the estimation of the mass based on the mass function of the system. Apart from that, one can only estimate the nature of the compact object by comparing its X-ray timing and spectral properties and X-ray-radio correlation with those of other known sources.

As we mentioned above, the energy spectra of LMXB systems are described by a thermal component and the Comptonised component. In addition the NS also show  emission from the surface of the NS and the so-called "boundary layer"; this component is generally described by a blackbody  \citep[e.g., ][]{Mitsuda1984, DiSalvo2000, Gierlinski2002, Lin2007}. It is also possible to use the presence of this additional component on NS the energy spectra to distinguish between BH and NS. Probably this is the most commonly used method when a new system is discovered, and at the same time it is probably one of the most unreliable methods. See, for example, the case of XTE J1812--182 \citep{Markwardt2008, intZand2017, Goodwin2019}, MAXI J1810--222 \citep{Maruyama2018, Negoro2019} or MAXI J1807+132 \citep{Shidatsu2017}. Following the detection of a new transient LMXB, the individual spectra obtained typically do not result in statistically significant deviations between the different spectral models in order to infer the nature of the corresponding compact object.

The identification  of the compact object can also be done based on the X-ray timing properties of the system. As we mentioned earlier, if coherent pulsations are found, we can determine unambiguously that the system hosts a NS \citep[see ][and references therein]{Patruno2012}. The presence of kilohertz quasi-periodic oscillations (QPOs) at frequencies between 300 Hz and 1200 Hz \citep[e.g., ][]{Klis2006, vanDoesburgh2018} strongly suggests that the compact object is a NS, too. However, the presence of the low-frequency QPOs in the mHz-50 Hz range does not always unambiguously pinpoint the nature of the system \citep{Klein-Wolt2008}. Both BH and NS are also similar in terms of broadband noise up to 500 Hz \citep{Klein-Wolt2008}. Above 500 Hz, the broadband noise of BH systems decreases while NS systems can show broadband noise up to higher frequencies \citep{Sunyaev2000}.
In terms of radio emission, BH systems are generally brighter than NS systems in the radio band, when observed at comparable X-ray luminosity  \citep{Fender2001, Fender2006, Migliari2006, Fender2014, Corbel2013}.

The nature of the compact object can also be identified by estimating the mass function of the system and measuring, estimating or assuming the mass of the companion star. The mass function only gives a lower limit of the mass of the compact object given uncertainties in the inclination of the system. 
If the compact object is $>4-5 M_\odot$, then it is usually agreed that system contains a black hole \citep[e.g., ][]{Casares1992, McClintock2001, Orosz2003, Casares2007, Darias2008}. 
If it is of the order of $2M_{\odot}$ or less, then it is most probably a neutron star \citep{Lattimer2004, Lattimer2007, Demorest2010, Lattimer2012, Orosz2003, Casares2007, Ziolkowski2008}.

In some rare occasions, the mass estimate is in the $2M_{\odot}<M<4M_{\odot}$ range. In this case, it is not possible to determine unambiguously the nature of the compact object.
GRO J0422+32 gives a good example of the limitations of this method. \cite{Gelino2003} estimated the mass of GRO J0422+32 to be $3.97\pm0.95$ $M_{\odot}$, and therefore a black hole identification. However, about 10 years later \cite{Kreidberg2012} explored possible systematic underestimations of the inclination of X-ray binary systems, which can increase the mass of the compact objects. They found this was the case of GRO J0422$+$32 and, taking into account this underestimation, they obtained a mass of 2.1 $M_{\odot}$, suggesting that GRS J0422$+$32 was instead a NS system.

However, it is not always possible to have an estimation for the mass of the companion star and, as a result, estimate the mass function of the system. Despite this fact, it is still possible to estimate the mass function of the compact object. \citet{Casares2015} found a correlation between the full width half maximum (FWHM) of the H$_{\alpha}$ line of the accretion disc and the velocity semi-amplitude of the companion star that, combined with supplementary information on orbital periods, can be used to estimate the mass function of the compact object from single epoch spectroscopy. Another correlation between the mass ratio of the binary system and the ratio of the double-peak separation to the line width can be used to estimate the mass function of the system \citep{Casares2016} and, from there, try to determine the nature of the compact object. 

All the methods of classifying LMXB sources that have been employed so far have had their own drawbacks. One technique that is yet to be explored to classify LMXBs is the use of machine learning algorithms. Machine Learning (ML) algorithms have been successfully used to solve problems in various domains of astronomy. They have been used to identify the furthest quasars in the universe \citep{Mortlock2011}, classify galaxies based on their morphology \citep{Storrie1992,Bazell2001,Calleja2004,Banerji2010}, to detect small near-earth asteroids \citep{Wasz2017} and even for hunting exoplanets \citep{,Thompson2015,Pearson2018}. 
Machine learning has also been applied in the X-ray domain by \citet{Huppenkothen2017} to classify light curves of the unusual BH X-ray binary GRS 1915+105. 
An effort to distinguish between different types of X-ray binaries has been reported  by \citet{Gopalan2015}, where they use a three-dimensional coordinate system comprising of colour-colour-Intensity diagrams to find clusters of data which can distinguish between BH and NS. 

In this work, we explore whether ML applied to the X-ray energy spectra of LMXBs can be used to identify the nature of the compact object. 
In order for ML algorithms to work, a large database of classified data is needed to develop a robust classification model. For this reason, we  use the full archive from the Rossi X-ray Timing Explorer (RXTE) mission \citep{bradt1993x}. This is probably the largest database today of X-ray observations of LMXBs, providing us with more than 8,500 observations from 33 NS systems and more than  6,000 observations from 28 BH systems.

The outline of the paper is as follows. Section~\ref{sec:data} describes the structure and composition of the data used in this work. In Section~\ref{sec:algo}, we explain the process of choosing a machine learning algorithm for classifying LMXBs along with a small description of the chosen algorithm - the random forest. This is followed by different methods employed and their results in Section~\ref{sec:methods}. In Section~\ref{sec:results} we analyze the results based on different factors which govern the classification and predict the classes for sources with unknown classification. Summary and future scope of this work is presented in Section~\ref{sec:summary}.

\section{Data Reduction and Preparation}\label{sec:data}

We used data from the Proportional Counter Array \cite[PCA;][]{Glasser1994} instrument aboard the Rossi X-ray Timing Explorer (RXTE). The PCA is an array of five proportional counters units (PCUs) with a total collecting area of 6500 cm$^2$; each PCU has an energy range of 2-60 keV. 
We selected a total of 61 sources which are classified as Black Hole (BH) or Neutron Star (NS) binaries, depending on the nature of the compact object. We chose those sources which have been extensively studied in the past and the classification is well known and consistent across different studies and catalogues \citep[see e.g., ][ for BH]{Corral-Santana2016,tetarenko2016watchdog}. After source selection, we obtain all data from pointed observations corresponding to these sources from RXTE archive \footnote{\url{https://heasarc.gsfc.nasa.gov/cgi-bin/W3Browse/w3browse.pl}}. 

To calculate X-ray colours we use the 16-s time-resolution Standard 2 mode data. For each of the five PCA detectors (PCUs) we calculate a soft and a hard colour, which are defined as the ratio between the count rate in the $6.0-16.0$ and the $2.0-6.0$ keV band, and the ratio between the $16.0-20.0$ and the $2.0-6.0$ keV band. We also calculate the intensity defined as the count rate in the $2-20$ keV band. To obtain the count rates in these exact energy ranges, we make a linear interpolation between all the PCU channels. We then carry out deadtime corrections, we subtract the background contribution in each band using the standard bright source background model for the PCA (version 2.1e1) and we remove instrumental drop-outs to obtain the colours and intensity for each time interval of 16s.  It is important to take into account that the \textit{RXTE} gain epoch changes with each new high voltage setting of the PCUs \citep{Jahoda2006}. We normalized our data to the Crab \citep[method introduced by ][]{Kuulkers1994} in order to correct for this effect and the differences in effective area between the PCUs.

For each observation we obtain the background, response and the spectrum files from which we extract the count-rate values of the desired energy spectrum range in a text file using the Xspec software \citep{Arnaud1996}.
We then reject all observations that have a net count-rate less than 5 counts per second in order to avoid low signal-to-noise spectra.

For each observation, we used 43 channels within the energy range of 5-25 keV. Machine learning algorithms require each observation to be of the same size hence we keep the number of channels fixed to 43.

We use these 43 count rate values directly as an input vector for the ML algorithm. 
 Due to variations in the sensitivity of particular channels with time, energy ranges tend to vary a little bit for each spectrum \citep{jahoda1996orbit,Jahoda2006}).
 The interstellar absorption $\textrm{N}_\textrm{H}$ can vary from source to source, and therefore adds another variable that the ML algorithm must take into account.  We found that in practice, ignoring data below the 5 keV range to avoid the effect of $\textrm{N}_\textrm{H}$ produced higher accuracy in classification. We chose the upper bound as above the energy value of 25 keV the instrument efficiency begins to deteriorate and the corresponding values contain minimal information about the spectrum. 
 We also choose to ignore other potential contributions or effects on the spectra as otherwise it would result in further reduction of our already small sample of spectra. Furthermore the objective of using machine learning is to identify intrinsic characteristics of the spectra belonging to two classes and accounting for these effects would add more human biases. Among the potential contributions and/or effects, we ignored that from possible absorption/emission lines on top of the X-ray continuum (i.e. the $\sim6.5$keV iron line). This is because such lines generally contribute only a few  percentage of the total flux, their strength varies differently between sources, between states of a given source, and in most cases are not resolved given the low-spectral resolution of the RXTE/PCA data. We also did not take into account the effects of the different source inclination. This  is because little is known about the inclination, and generally the uncertainties are very large \citep[][]{Munoz-Darias2013, Motta2015}.

\begin{figure*}
\centering

\includegraphics[width=1.0\textwidth]{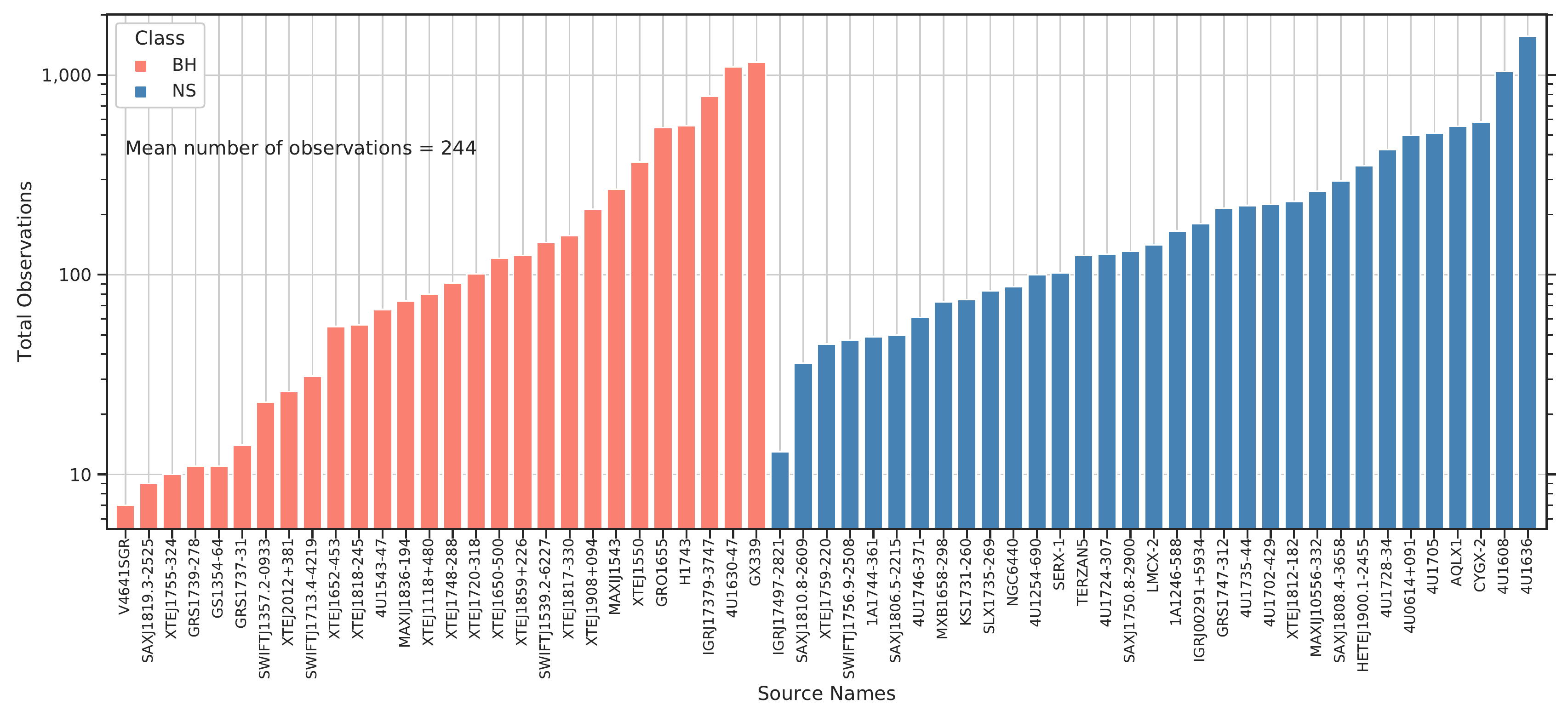}
\caption{Source wise distribution of data on black hole (BH) and neutron star (NS) classes. The mean is approximately 244 observations with some sources having 1000+ observations while others having less than 20. The list of individual sources along with their total number of observations and class labels can be also found in Table~\ref{tab:acctable2}}.
\label{fig:srs}
\end{figure*}

In the final dataset, we have a fairly balanced representation of the two classes with 8669 observations from 33 sources identified as neutron-star LMXBs (58\%) and 6216 observations from 28 sources identified as black-hole LMXBs (42\%). 
In Fig.~\ref{fig:srs} we show the number of observation per source for each class in the dataset. As can be observed, a few sources have \textgreater1000 observations while some have \textless20 observations.

\section{Algorithm selection and description}\label{sec:algo}

Machine learning is a branch of computer science that consists of algorithms which can learn to identify patterns in the data without any prior specification of a rule or model. By learning from the information in the data, a machine learning algorithm tries to approximate an underlying model that can define the data. Such models are used for handling various problems like classification, regression, clustering, etc. An algorithm tries to approximate these models in its ``training phase'' and based on the process it uses to approximate these models it is divided into two categories: supervised and unsupervised techniques.

For implementing a ML method, the dataset should contain a specific number of features for each input object. In the supervised training method, each set of input features corresponds to a label or a target value. The dataset is divided into train, validation and test sets. The  model is trained using the former, and then validated using the validation set. Multiple models with different initial settings are trained on the first set, and the best one is selected using the validation set. Supervised machine learning problems can further be divided into two types -  Classification and Regression. Simply put when the expected outcome is a real-valued number, it is considered a regression problem \citep{firth2003estimating, ramirez2001prediction, nesseris2012new} whereas when the objective is to categorize data, it is known as a classification problem \citep{Bazell2001, mcglynn2004automated, ball2006robust, zhao2007automated}.
In the unsupervised type of ML techniques, there is no requirement for a predefined label/class, and the algorithm tries to understand the relation between the input features without the help of the user. Some common examples of unsupervised learning include clustering tasks \citep{feitzinger1987fractal, wagstaff2005making, rebbapragada2009finding}, dimensionality reduction \citep{hojnacki2007x}, estimating the density function \citep{ferdosi2011comparison} and association.

In this work, we approach the problem of classifying an X-ray spectrum into either a BH or a NS. This is a supervised binary classification problem. 
There are several ML algorithms that can be used for handling this type of binary classification problem. As per the ``no free lunch theorem for Optimization'' \citep{Wolpert1997}, there is no one particular algorithm that excels in all scenarios. However, there are a few points the user should consider while selecting the right machine learning algorithm. In our case,  the first criterion is accuracy. The algorithm which can provide the highest percentage of correct classifications is usually the most favorable. 

One of the weaknesses of using machine learning methods is that they are a ``black-box'' when the user wants to understand the decision-making process that lead to a given result.
This property of a machine learning algorithm is known as interpretability. Sometimes the most accurate algorithms are the least interpretable, or vice-versa. Therefore, there is usually a trade-off between the two criteria for the selecting the best algorithm \citep{nakhaeizadeh1997development}. 
It is worth mentioning that it is possible to study the decision-making process of an algorithm; however, the nature of the data can make it very difficult (or virtually impossible) to understand the process.
In cases where the data have features (or input vector to the machine learning algorithm) that have some direct physical meaning (for example temperature, mass, etc.), it is possible to draw correlations or understand which physical feature has the most significant contribution to the decision-making process.
In the problem studied in this paper,  data consist of count-rate values corresponding to a certain energy range. Therefore it is very difficult to visualize and/or understand  the decision-making process. Therefore, we decided that it was more favorable to choose an algorithm that is more accurate, even if it  compromised  the interpretability. 

In this work we experimented with the following algorithms:

\begin{itemize}
    \item \textit{\textbf{Classification and Regression Trees (CART) or more commonly known as Decision Trees}} \citep{breiman1984classification}: use a tree like structure to map the input vector to the target values. Based on the target values they can be either classification trees or regression trees. 
    
    \item \textit{\textbf{Random Forest (RF)}} \citep{Breiman2001}: is an ensemble method that combines the output of several decision trees to improve on the prediction of a single tree. As we will see in sub-section \ref{comparison} this method has the highest accuracy compared to the other algorithms and therefore is our algorithm of choice. We will talk about it in more detail later in section \ref{sec:rf}.
    
    \item \textit{\textbf{XGBoost (XGB)}} \citep{Chen2016}: is another ensemble method that implements machine learning algorithms in a gradient boosting framework \citep{mason2000boosting} to improve efficiency and speed.
    
    \item \textit{\textbf{Logistic Regression (LR)}} \citep{Cox1958}: is a multivariate analysis model that predicts the probability of membership to any class based on the values of some predictor variables; these variables are not constrained to follow a given (normal) distribution, not even be continuous.

    \item \textit{\textbf{k-Nearest Neighbors (KNN)}} \citep{Cover2006}: is a nonparametric classification technique that works on the following simple principle: Given a query for prediction, it finds the k closest neighbors to the data point in the training sample by calculating the euclidean distance from every point and then assigns the class which is the most common amongst its k nearest neighbours.
    
    \item \textit{\textbf{Support Vector Machines (SVM)}} \citep{Cortes1995}: is a type of kernel-based algorithm that builds a set of hyperplanes in the high-dimensional feature space such that they have the maximum possible distance from the nearest data point of any class thus optimizing the separation between the different classes in the data.
\end{itemize}

For further reference on the detailed workings of these algorithms see \cite{Ivezic2014}, an astronomy-oriented textbook for Machine learning.

We chose these algorithms as they fall into the category of traditional machine learning algorithms that are usually known to show satisfactory performance even with a limited amount of data. They also have significantly lower execution times as compared to the widely popular deep learning methods (see, for e.g., \citet{kotsiantis2007supervised}). 

\subsubsection{Random Forest}\label{sec:rf}

Random forest is an ensemble technique which is used to boost the prediction made by an individual decision tree \citep{Breiman2001}. A decision tree is one of the most intuitive yet powerful machine learning algorithms \citep{breiman1984classification}. A decision tree is made up of branches of nodes, where sets of if-this-then-that rules are applied to the features of the input data, and based on the result, lead down one of the branches of the tree. The final layer of nodes, also known as leaf nodes, contains a predicted class label which is compared to the expected class for a particular input vector. Although the decision tree algorithm has proven to be very efficient (see, for e.g., \cite{vasconcellos2011decision}), a decision tree, if improperly trained, can at times over-fit the data \cite[Chapter~3]{mitchell1997machine}. The idea behind random forest is to combine the decisions of several such trees to improve upon the decision of a single over-trained tree. Taking a majority-vote over the decision of all the trees helps in reducing the variance of the predictions \citep{Breiman2001}. 
The probability of a source belonging to one class or the other is also calculated in a similar way, i.e., by dividing the number of trees that predicted the same class by the total number of trees.
The basic working of a random forest algorithm is explained below:

\begin{enumerate}[label=\arabic*.,leftmargin=1\parindent,align=left]

    \item From a total number of $K$ input features, the algorithm chooses a number $I$ such that $I \ll K$.

    \item Using bootstrap sampling, the algorithm chooses a training set for a tree by selecting a subset from the complete training data. It keeps the remaining data for validating the predictions.

    \item The algorithm chooses $I$ random features at every node of the tree and then calculates the most optimal split for the training set using these features.

    \item The algorithm grows every tree to its maximum depth without any pruning (unlike a solitary decision tree which is pruned after growing fully to prevent overfitting \citep[see][]{breiman1984classification}.

    \item The algorithm then repeats the above step to generate many such trees.

    \item After the training is completed, the algorithm uses a majority vote to predict the  class of the input data. To calculate the majority vote for a given input vector, the algorithm selects the class which was predicted by the majority of individual trees. To calculate the probability/confidence of the prediction the algorithm uses the ratio of trees that predicted the particular class to the total number of trees. 

\end{enumerate}

A decision tree algorithm works from top to bottom (see Fig.~\ref{fig:rf}) and usually chooses a variable at each step that optimally splits the set. Depending on the particular splitting algorithm used, the selection process of the variable varies. Here we use the default gini impurity method \citep{breiman1984classification} which is a measure of the likelihood of an incorrect classification of a randomly chosen element, if the element was randomly labeled according to the distribution of labels in the dataset.

We illustrate the decision making process of a random forest algorithm in Fig.~\ref{fig:rf}. We implement the random forest algorithm using the \texttt{scikit-learn}\footnote{\url{https://scikit-learn.org/stable/}} \citep{skl} library of python.
We use grid search combined with cross-validation to find the best hyper-parameters for the algorithm. Hyper-parameters are a set of parameters defined prior to the training process that are used to tune the performance of the ML algorithm. The optimal hyper-parameters obtained were:

\begin{figure}
    \includegraphics[width=\columnwidth]{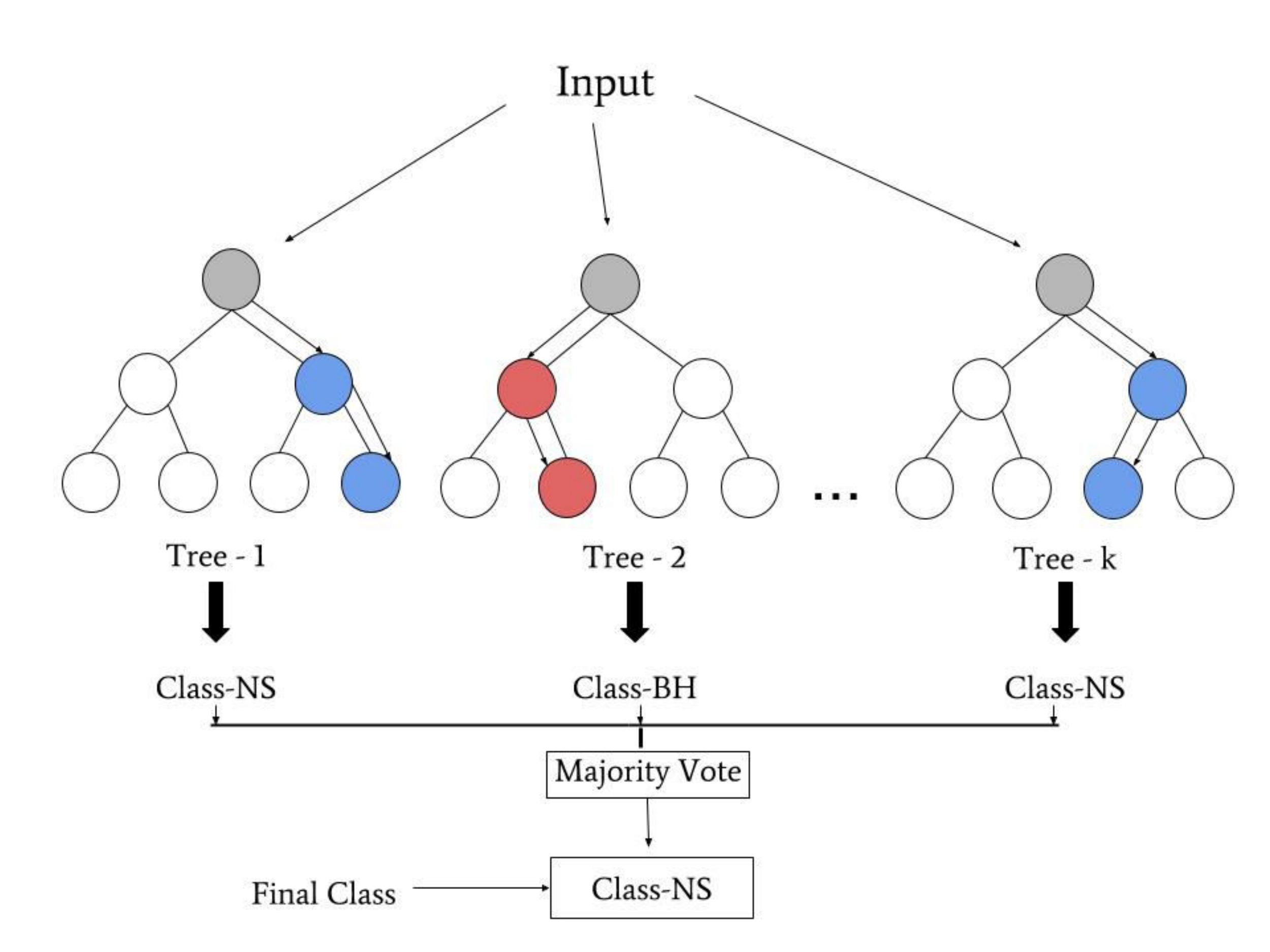}
    \caption{Illustration of the decision making procedure in a Random forest algorithm.}
    \label{fig:rf}
\end{figure}

\begin{itemize}[leftmargin=1\parindent, align=left]
    \item[--] Min\textunderscore samples\textunderscore leaf\,=\,3 (The minimum number of samples required to be at the leaf node)
    \item[--] Min\textunderscore samples\textunderscore split =\,8 (The minimum number of samples required to split an internal node)
    \item[--] Max\textunderscore features\,=\,2 (The number of features to consider when looking for the best split)
    \item[--] N\textunderscore estimators\,=\,1000 (The number of trees in the forest)
\end{itemize}

\subsection{Comparison of algorithms} \label{comparison}

For selecting the best classification method, we train and test different algorithms and compare them using `accuracy' as a metric, which is defined as the ratio of the number of observations correctly classified to their class (NS or BH) to the total number of observations. 

\begin{figure}
    \includegraphics[width=\columnwidth]{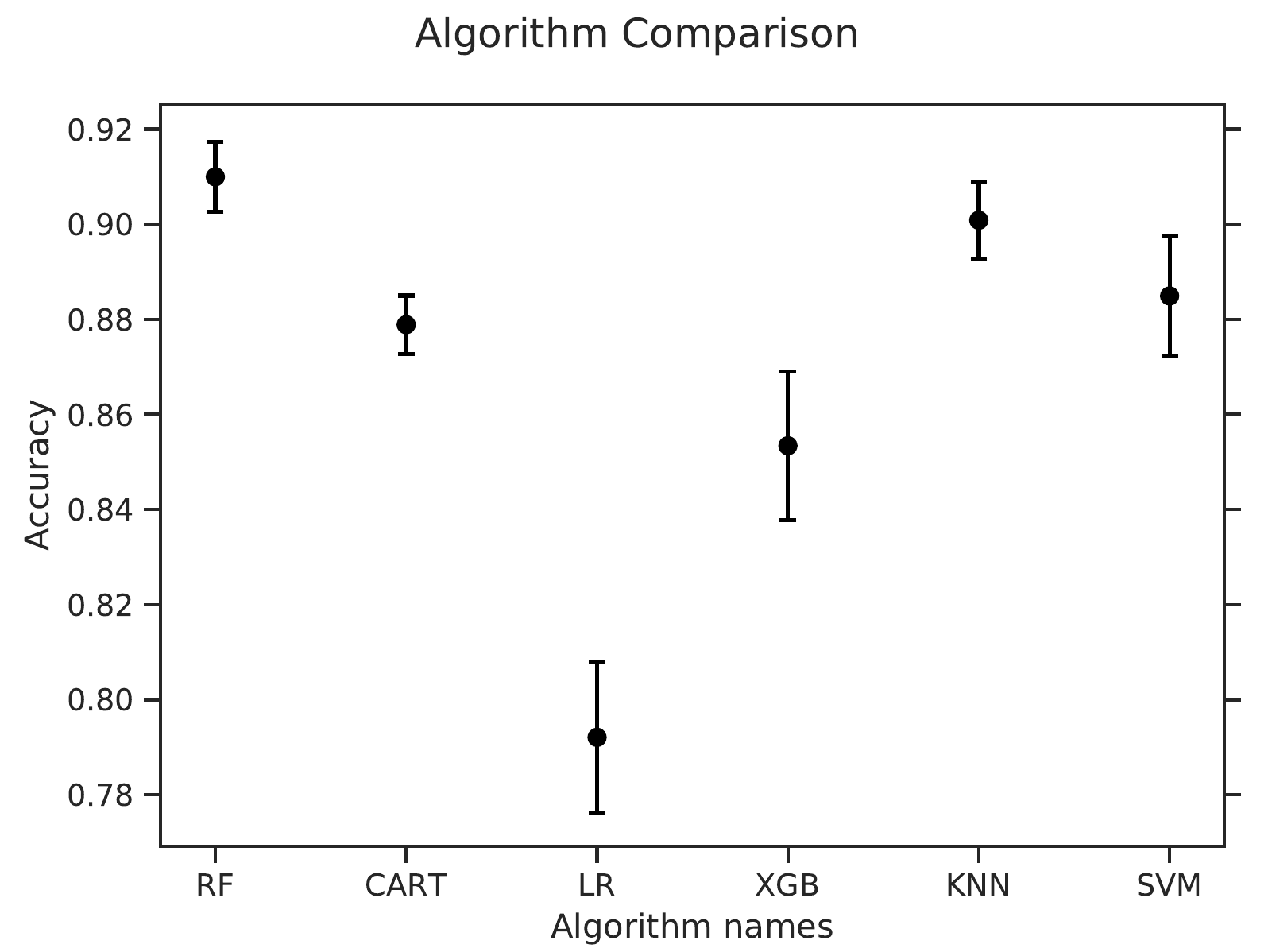}
    \caption{Comparison of the performance of different machine learning algorithms using the 10-fold cross-validation process. The algorithms on the x-axis (from left-to-right) are Random Forest, Decision Tree (CART), Logistic Regression, XG-Boost, K-Nearest Neighbors and Support Vector Machines. The Random Forest performs the best with an overall accuracy of $91\pm 2$\%.}
    \label{fig:algo-comp}
\end{figure}

To compare the algorithm, we first split  the dataset consisting of 14885 observations into training and test sets and use k-fold cross-validation technique \citep{burman1989comparative}, in which we divide the dataset into k even samples. Then we use one sample as a test set while training on the remaining k-1 samples. We repeat this process for each of the k samples in the process covering the entire dataset. We use 10-fold cross-validation (k=10) along with the default hyper parameters for each algorithm. Results of the 10-fold cross-validation and comparison between the different algorithms are presented in Fig.~\ref{fig:algo-comp}. We find that the random forest algorithm performs the best among all the selected methods, giving the highest accuracy of $91\pm 2$\%. 
Therefore, in the following sections we only report on the results of the classifications obtained using the random forest algorithm.

\section{Methods and Initial Results}\label{sec:methods}

We apply the RF algorithm with the best combination of hyper-parameters to the dataset described in Sec.~\ref{sec:data}. Since the dataset contains 14885 observations for 61 individual X-ray sources, each source is represented by multiple observations taken at different times. 
Given that the LMXBs studied here are variable in nature,  different observations for the same source could be sampling a different physical spectral state (i.e. different geometrical configuration). 
Therefore the classification of the energy spectra of LMXBs can be treated as any other typical ML binary classification problem where each observation is considered independent of the others. %
However, due to the nature of the problem and the limitations of our data (e.g., time-variability factors in the data, correlations between spectra of the same source taken at different times, and unequal number of observations for different sources), we had to use different strategies to train the model and evaluate its performance. We used: 

\begin{itemize}

\item [i] Traditional train-test split (Observation-wise splitting): In this approach, the 14885 spectra are randomly split into a training and test set consisting, respectively, of 90 percent and 10 percent of the observations. Here, we assume that each observation is independent of the rest, meaning that there are no correlations between different observations for the same source.

\item [ii] Source-wise splitting: Rather than splitting on the basis of observations, we split the dataset into training and test sets on the basis of sources. We use spectra corresponding to 34 sources for the training and the testing is performed on the remaining 27 sources.

\item [iii] Leave-one source out: In this method, we train the RF model on all observations corresponding to all sources except one. The observations corresponding to the excluded source are used for the testing.

\end{itemize}

Detailed description of each of these approaches is provided in the following subsections.

\subsection{Method 1: Traditional train-test split}\label{sec:method1}

The orthodox way to perform any machine learning classification experiment is to divide the complete dataset into train and test sets. For this, we used the \texttt{train\textunderscore test\textunderscore split} function of the \texttt{scikit-learn} \citep{skl} python library. We keep 90 percent of the data (13396 observations) for the training and validation. The remaining 10 percent of the data (1500 observations) are used for the testing. 
We train the RF algorithm with the best combination of parameters described in Sec.~\ref{sec:rf}. %
Testing the trained RF model results in an overall accuracy of $\sim$ 91\%. Table~\ref{tab:tr-tt} shows the performance of the classifier for observations of both classes. The performance is equally good for the two classes. 

\begin{table}
  \centering
  \begin{tabular}{ccccc}
  	\hline
    Class  & No. of   & Correctly  & Misclassified & Accuracy\\
           & test obs & Classified &               &     (\%)\\
    \hline
    NS & 867 & 814 & 53 & 94 \\
    BH & 622 & 545 & 77 & 88 \\
    \hline
    Total & 1489 & 1359 & 130 & 91 \\
    \hline
  \end{tabular}
  \caption{Performance of the algorithm for the two classes using the traditional train-test split method.} 
  \label{tab:tr-tt}
\end{table}

The major drawback with the traditional train-test split method for our case is that it does not take into account potential correlations between different spectra of a given source. As a result, some observations from the same source might be used in both the training as well as test set.
Testing the classifier on different observations of a source which also had some of its data in the training set could lead to a biased and overestimated value of the accuracy as the classifier would be able to identify spectra belonging to the same source very easily. However in the real-life scenario we would have data from a newly discovered X-ray source that needs to be classified. Since it is not possible to determine the expected accuracy for the real-life scenario with this method, we only use it for comparing the performance of different algorithms and choosing the best amongst them (Sec.~\ref{sec:algo}).

\subsection{Method 2: Source-wise train-test split}\label{sec:method2}

\begin{figure}
    \includegraphics[scale=0.5]{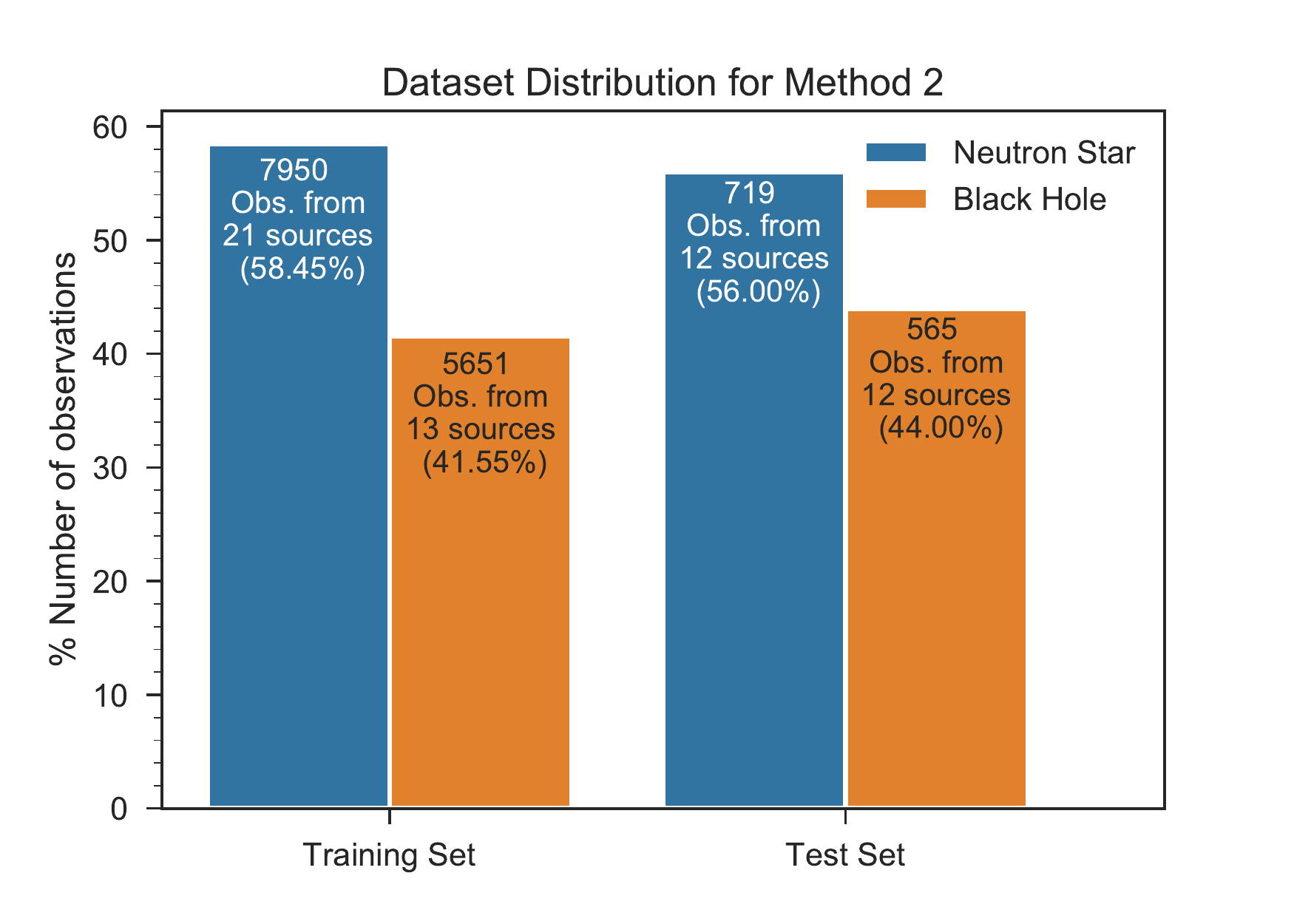}
    \caption{Distribution of the data for the source-wise train-test split method. Both the train and test datasets have a good ratio of data for the two classes.} 
    \label{fig:tdtdb}
\end{figure}

To avoid the shortcomings of the traditional observation-wise train test split, we split our data source-wise, i.e., we select some sources to be used for training, while testing on the remaining sources. In order to maximize the usage of data available for training, we choose all the sources with less than 100 observations as the test sources while the remaining are used for training. With this criterion, we had a training set of 34 sources with a total of 13,601 observations ($\sim$ 90\% of the data) and a test set set of 27 sources with a total of 1284 observations ($\sim$ 10\% of the data). The training set consists of 7950 BH LMXB observations from 21 sources (58\%) and 5651 NS LMXB observations from 13 sources (41\%). The test set consists of 719 BH LMXB observations from 15 sources (56\%) and  565 NS LMXB observations from 12 sources (44\%). These details are also represented in a graphical form in Fig. ~\ref{fig:tdtdb}.

As can be observed from the figure, a satisfactory ratio between BH and NS observations is maintained in the train and test sets. The complete list of 27 sources used in the test set, actual class of each source from the literature and total number of observations for each source are listed in Table~\ref{tab:acctable1}. For each source, we provide all observations corresponding to that source to the classifier and each observation is assigned to `BH' or `NS' class. The percentage accuracy is computed by dividing the number of observations assigned to the actual class by the total number of observations for that source. We provide the percentage accuracy for each source in the test set in the last column of the table. 

\begin{figure*}
    \includegraphics[width=\textwidth]{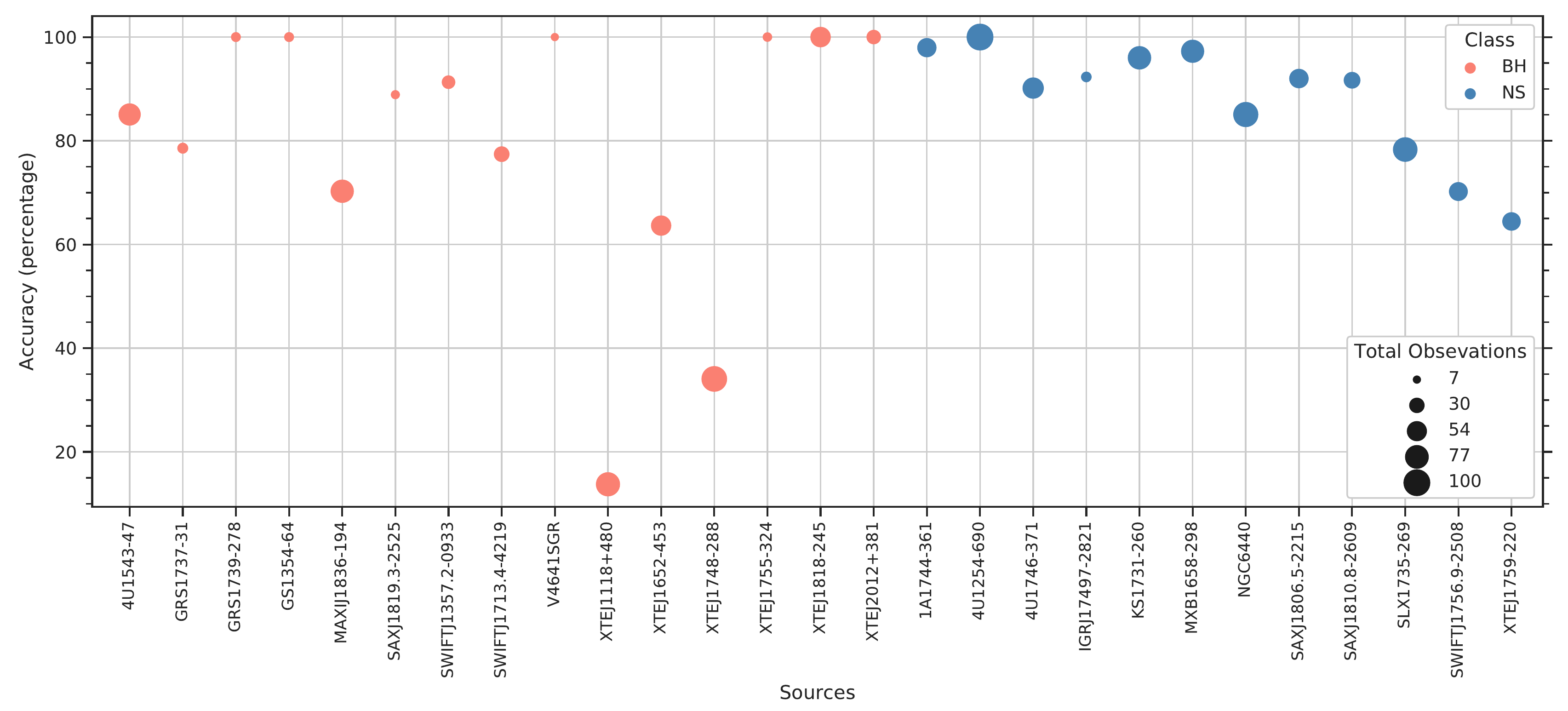}
    \caption{Plot showing individual source wise accuracies for sources in the test set of the source-wise train test split method. The points are coloured based on the classes and the size of the points correspond to the number of observations in each source.} 
    \label{fig:accuracy1}
\end{figure*}

The results obtained with this approach are also presented in Fig.~\ref{fig:accuracy1} where most of the sources have above 60\% accuracy and only two sources have less than 50\% accuracy. For a quantitative analysis of the algorithm's performance, we calculate the sigma-clipped average source-wise accuracy using \texttt{sigma\_clipped\_stats} function of the astropy python library \citep{astropy}. The sigma-clipped average accuracy gives an outlier resistant estimate of the algorithm's performance where the points lying beyond 3$\sigma$ from the mean value are iteratively removed while computing the statistic. Sigma-clipped mean percentage accuracy for the test set comes out to be 88\% with a standard deviation of 12\%. Further class-wise performance is detailed in Table~\ref{tab:sr-tt}. 

\begin{table}
  \centering
  \begin{tabular}{cccc}
  	\hline
    Class  & No. of Sources & Avg. \% Accuracy & $\sigma$ \\
    \hline
    NS & 12 & 88 & 11 \\
    BH & 15 & 80 & 25 \\
    \hline
    Total & 27 & 88 & 12 \\
   	\hline
  \end{tabular}
  \caption{Performance of the RF classification model for NS-BH classification using the source wise train-test split method.}
  \label{tab:sr-tt}
\end{table}

\subsection{Method 3: Leave-one source out}\label{sec:method3}

The source-wise train-test split method discussed in Sec.~\ref{sec:method2} closely mirrors the scenario we may have in terms of the available number of observations for a new source (which is not likely to exceed 100). However, the major drawback with the source-wise train-test split approach is that the test set remains unutilized for the training of the model. Although the test set contains only $\sim$10\% of the observations, these observations might occupy a region in the model space crucial for identifying the classification boundary (the boundary in the model space that separates the data of the two classes) which might not be represented  by the observations in the training set.
Therefore, in order to optimize the usage of available data, we use the leave-one source out method, where we keep all observations from one source as our test data while using all the remaining sources for training. We repeat this experiment for each source, so that we have the results for all the observations from each of the 61 sources. 

\begin{figure*}
    \includegraphics[width=\textwidth]{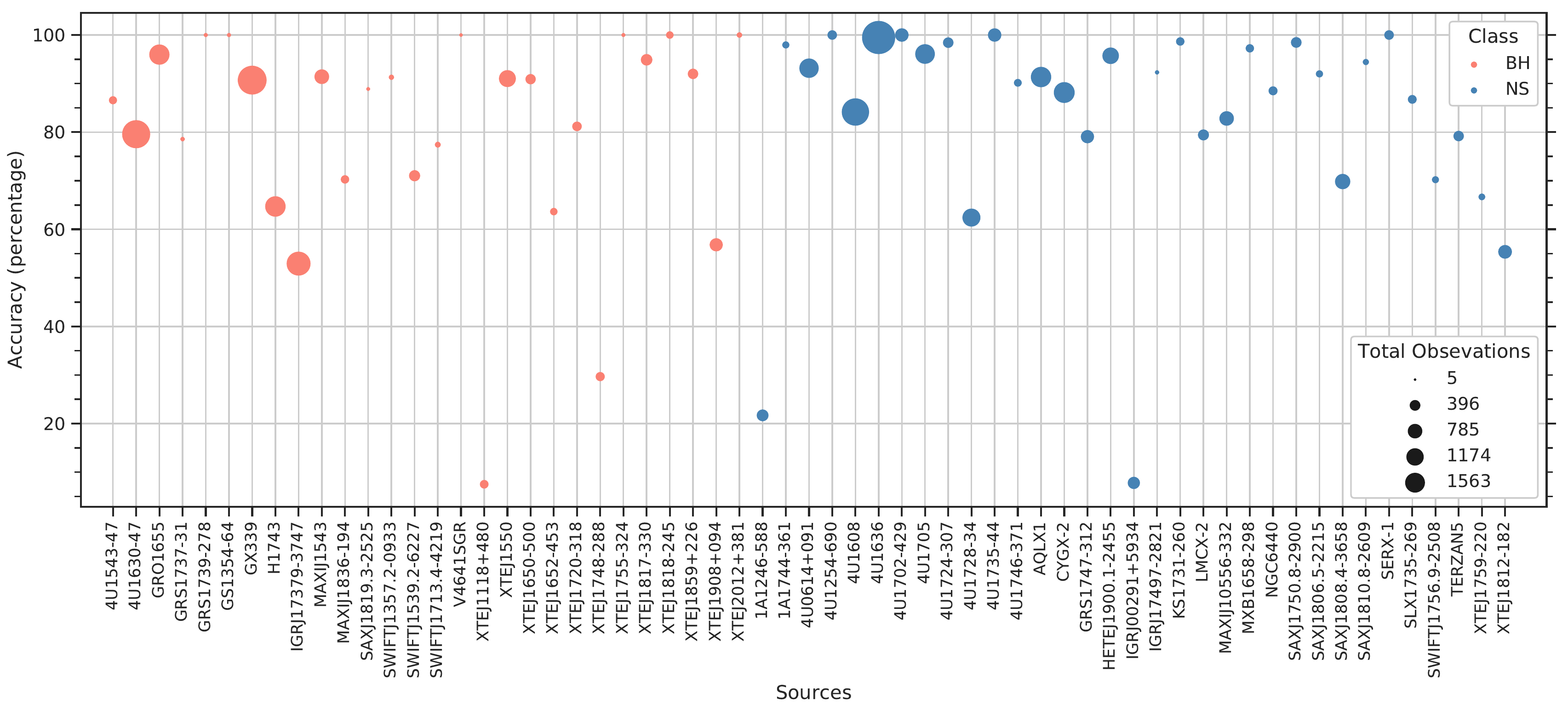}
    \caption{Plot showing individual source wise accuracies using the leave-one source out method of cross-validation. The points are coloured based on the classes and the area of the points corresponds to the number of observations in each source.} 
    \label{fig:accuracy2}
\end{figure*}

In the leave-one-source-out method,  the size of the training and test sets vary in each run. Our final model would be trained on the entire dataset whereas in this method, each model is using one source less than what the final model would use. 
Therefore, the coverage of the model in the feature space through this approach is closest to the final model. This can also be seen as a type of cross-validation method tailored for our data. We present the resulting accuracy for each source calculated through this method in Table~\ref{tab:acctable2} and in Fig.~\ref{fig:accuracy2}.

There are four sources that lie below the 50\% average accuracy mark. The sigma-clipped average accuracy using this method comes out to be $87\pm13\%$, which gives a lower bound proxy on the performance of our final model. We present the class-wise performance in Table~\ref{tab:los}.

\begin{table}
  \centering
  	\begin{tabular}{cccc}
  	\hline
    Class  & No. of Sources & Avg. Accuracy & $\sigma$ \\
    \hline
    NS & 33 & 89 & 11 \\
    BH & 28 & 85 & 14 \\
    \hline
    Combined & 61 & 86.63 & 13.08 \\
   	\hline
  \end{tabular}
  \caption{Performance of the RF classifier for NS-BH classification using the leave-one source out method. Average accuracies and standard deviations are computed using 3$\sigma$ clipping to get a robust estimate of the statistics.}
  \label{tab:los}
\end{table}

\section{Results and interpretation}\label{sec:results}

The average accuracies of the sources for both method 2 and method 3 are similar but less than that of method 1 (91\%). This is expected because of the bias in method 1 discussed earlier in section ~\ref{sec:method1}.
While the average accuracy decreases for methods 2 and 3 when compared to method 1, it is safe to say that the machine learning algorithm seems to do a satisfactory job in the overall classification of low mass X-ray binary sources. The lower bound of the accuracy ($87\pm13\%$) indicates that the Random Forest (RF) algorithm is able to identify the classification boundary between the two types of X-ray sources in the 43-dimensional space of their energy spectra. However, we note that there are a few sources for which the accuracy is very low and most of the observations of those sources are misclassified. In particular there are four sources, namely  XTE J1118+480 (BH), XTE J1748--288 (BH), IGR J00291+5934 (NS), and 1A 1246--588 (NS), which have less than 50\% accuracy out of which the observations of XTE J1118+480 and XTE J1748--288 are consistently misclassified with overall accuracy percentage of $\sim$ 10\% and $\sim$30\%, respectively, in methods 2 and 3. This motivates us to study these sources in more detail and probe the possible reasons for the misclassification of their spectra. It is difficult to determine the reasons for the misclassifications directly from the RF algorithm. Therefore we study the correlations between predictions of the RF algorithm and the factors that can influence them. Two such factors that can influence the energy spectra are the Signal-to-Noise Ratio (SNR) and the physical states of LMXB systems.

\subsection{Effect of Signal-to-Noise Ratio (SNR)}\label{sec:snr_analysis}

\begin{figure}
    \includegraphics[width=\columnwidth]{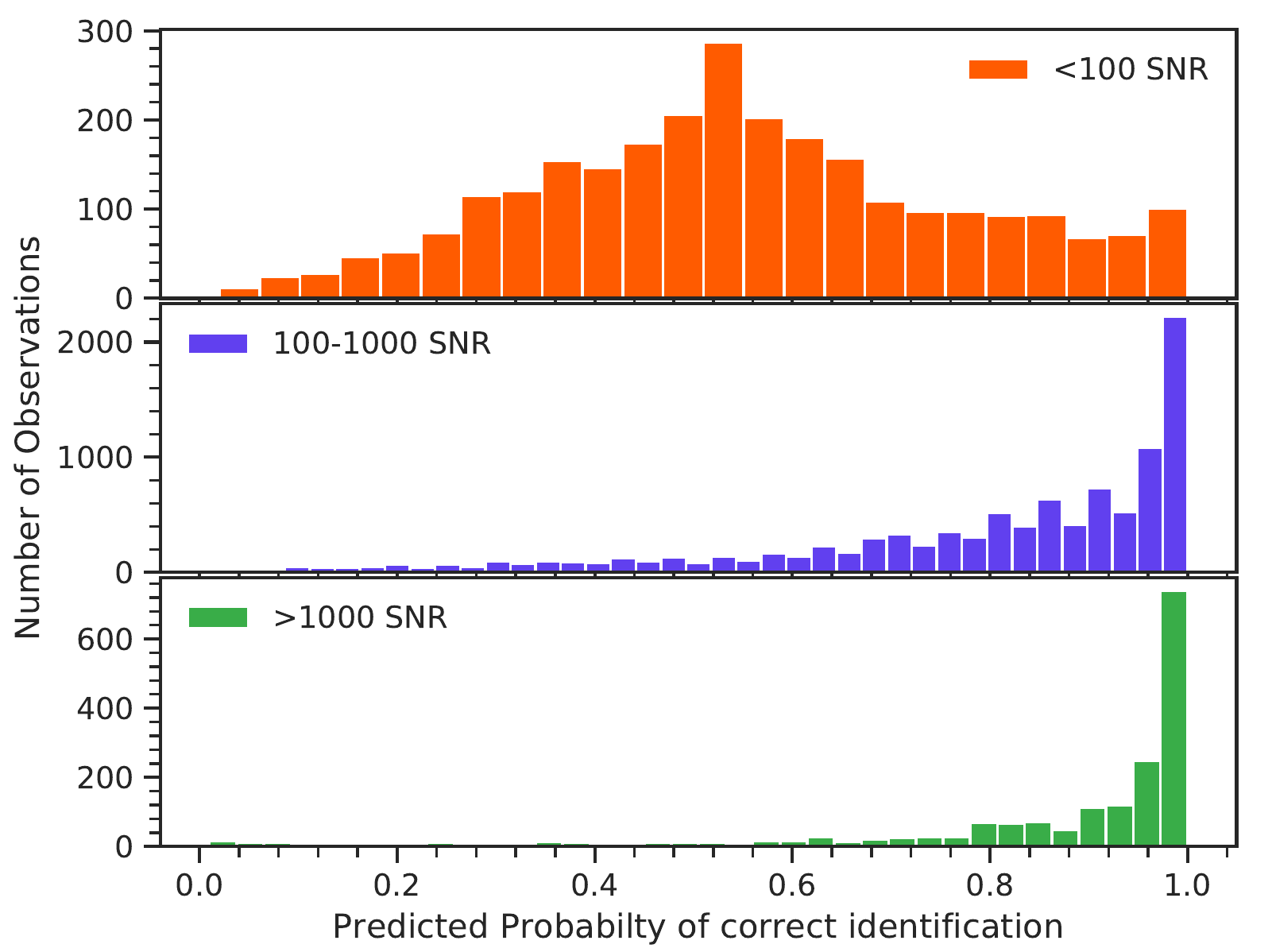}
    \caption{Distribution of predicted probabilities (confidence) for different SNR ranges. The distributions peak at 0.58, 0.87 and 0.91 (top to bottom), indicating that observations with higher SNR are correctly classified with a greater confidence.}
    \label{fig:snrplt}
\end{figure}

For each observation SNR is calculated by dividing the net count rate by the error in the net count rate. This information is obtained from the header of the spectra.pha file of the observations retrieved from the RXTE archive for each source. The SNR ranges from  as  low  as  4  to  more  than  5800. To investigate the influence of SNR on the classification, we divide all observations in three SNR ranges, <100, 100-1000, >1000, and analyze the predicted probability of classification using Method 3 (Sec~\ref{sec:method3}). The predicted probabilities are obtained using the \texttt{predict\_proba} function of the Random Forest model and serve as a measure of classification confidence. The distribution of predicted probabilities of correct identification for all observations in different SNR ranges is shown in Fig.\ref{fig:snrplt}. For observations with SNR <100, the distribution of predicted probabilities peaks at 0.58. For the other two SNR ranges, 100\,-\,1000 and >1000, the distribution peaks at 0.87 and 0.91, respectively. These results are also presented in Table~\ref{tab:snr}. This analysis indicates that the performance of the classification model increases with the increase in SNR. 

\begin{table}
  \centering
  \begin{tabular}{cccc}
  	\hline
    SNR Range  & Mode of Predicted & Total Obs. & Data (\%) \\
    & Probabilities & & \\
    \hline
    \textless 100 & 0.58 & 2706 & 18.2 \\
    100-1000 & 0.87 & 10348 & 69.5 \\
    \textgreater 1000 & 0.91 & 1831 & 12.3 \\
   	\hline
  \end{tabular}
  \caption{Predicted probability and distribution of observations for different ranges of SNR.} 
  \label{tab:snr}
\end{table}

We further investigated the misclassified sources by checking their average SNR. Among all the sources, only 1A1246-588 had an average SNR less than 100 (avg. SNR\,=\,48). This analysis indicates that, while the accuracy of the prediction increases in general with increasing SNR, a low SNR alone is the main reason behind the poor classification of the spectra for some sources.

\subsection{Correlation between predicted probability of correct identifications and state transitions}

\begin{figure*}
\hspace*{-1cm}
\subfloat[Sample sources with $>$50 \% accuracy\label{fig:cci1}]{%
\includegraphics[width=0.55\textwidth]{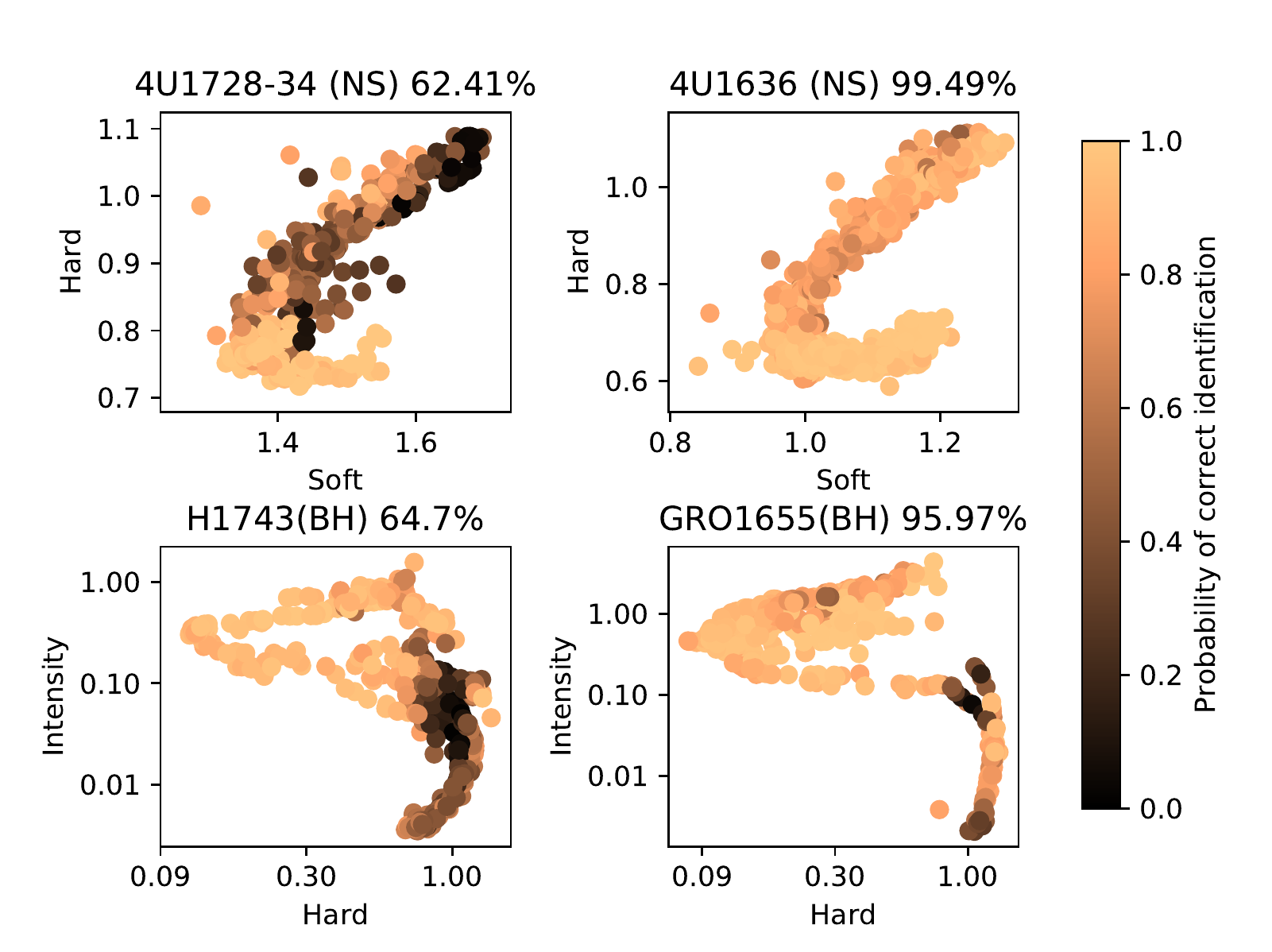}
}
\subfloat[Sample sources with $<$50 \% accuracy\label{fig:cci2}]{%
\includegraphics[width=0.55\textwidth]{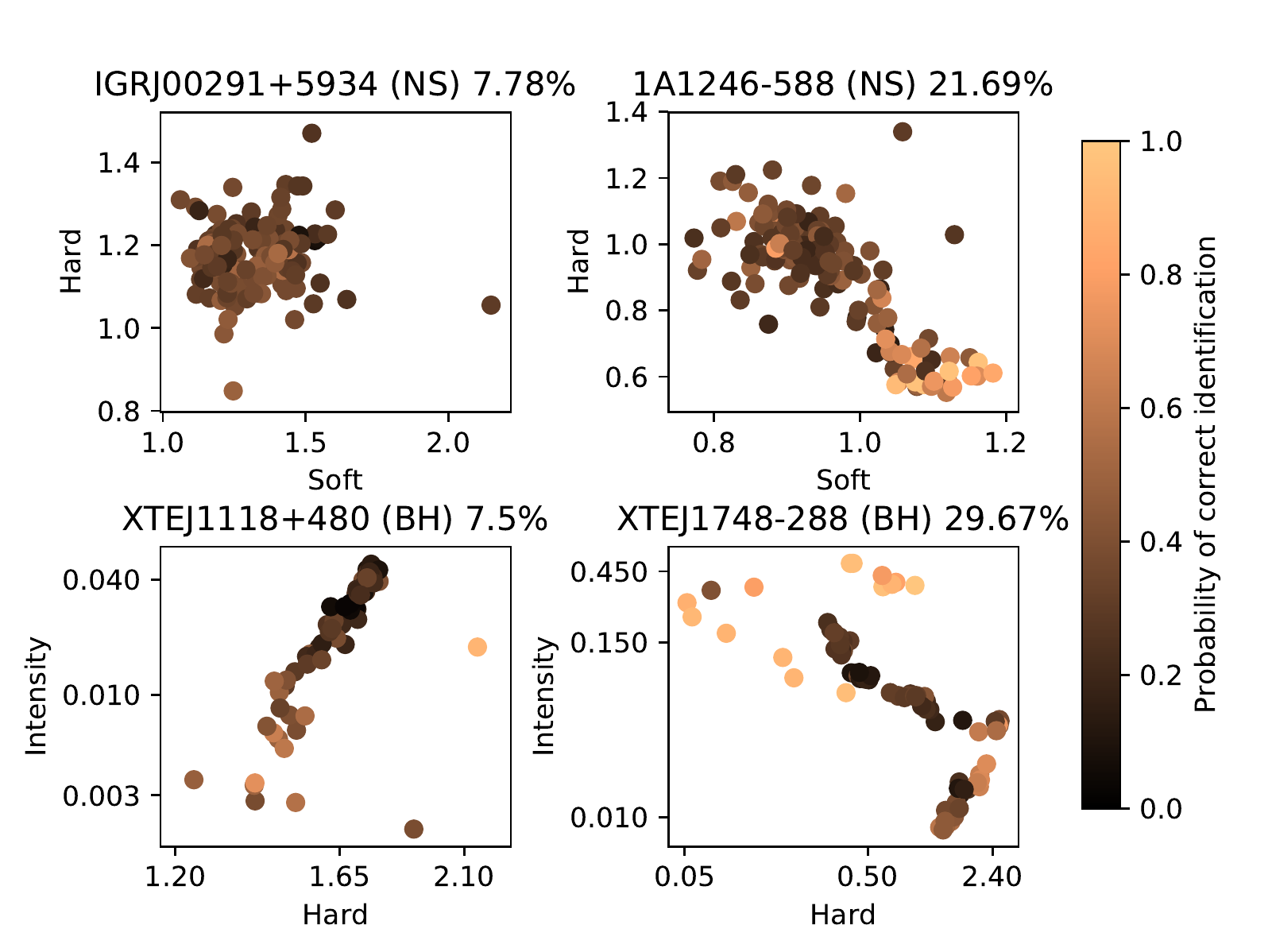}
}
\caption{Colour-Colour Diagrams (CCD) and Hardness Intensity Diagrams (HID) for NS and BH LMXB systems respectively. Left panel shows the CCDs and HIDs for two NS (top) and two BH (bottom) sources with good classification accuracy. Right panels show the same diagrams for poorly classified NS and BH systems using the RF classifier. Probability of correct identification for each observation using Method 3 is colour coded as shown in the adjacent colour bars. The source identifiers and their original classes are indicated on top of each diagram with their accuracy percentage from Table~\ref{tab:acctable2}.} 

\label{fig:cci}
\end{figure*}

In Fig.\ref{fig:cci1} we plot the CCD diagrams of two atoll-NS LMXBs (top panel) and HID diagrams of two BH LMXBs (bottom panel). The two atoll-NS sources are 4U 1728--34 (423 observations) and 4U 1636--53 (1563 observations) and the two BH sources are H 1743--32 (558 observations) and GRO J1655--40 (546 observations). We chose these systems as they have observations sampling all the typical spectral states. 

We colour each observation based on the predicted probability of correct identification obtained from Method 3 (leave-one source out) as shown in the colour bar plotted on the right side. 
Most of the misclassified observations (darker coloured circles) belong to the LHS or intermediate states while the HSS observations are very well classified (lighter coloured points). 

In Fig.~\ref{fig:cci2} we show the HID and CCD diagrams for the  4  sources for which our algorithm performs the worst. While the state transitions for these sources are not as pronounced as for the sources shown in fig.~\ref{fig:cci1}, we can still observe that the misclassifications (dark points) are predominantly in the hard region of the spectra.

\begin{figure*}
\subfloat[NS\label{fig:hcvp1}]{%
\includegraphics[width=\textwidth]{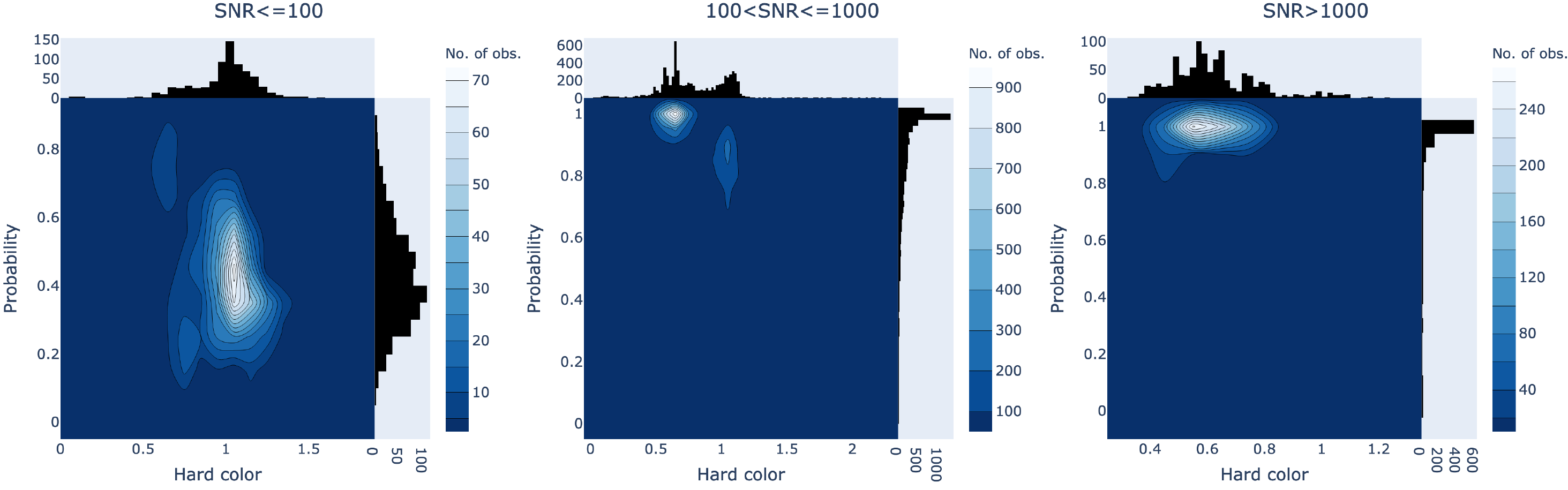}
}
\newline
\subfloat[BH\label{fig:hcvp2}]{%
\includegraphics[width=\textwidth]{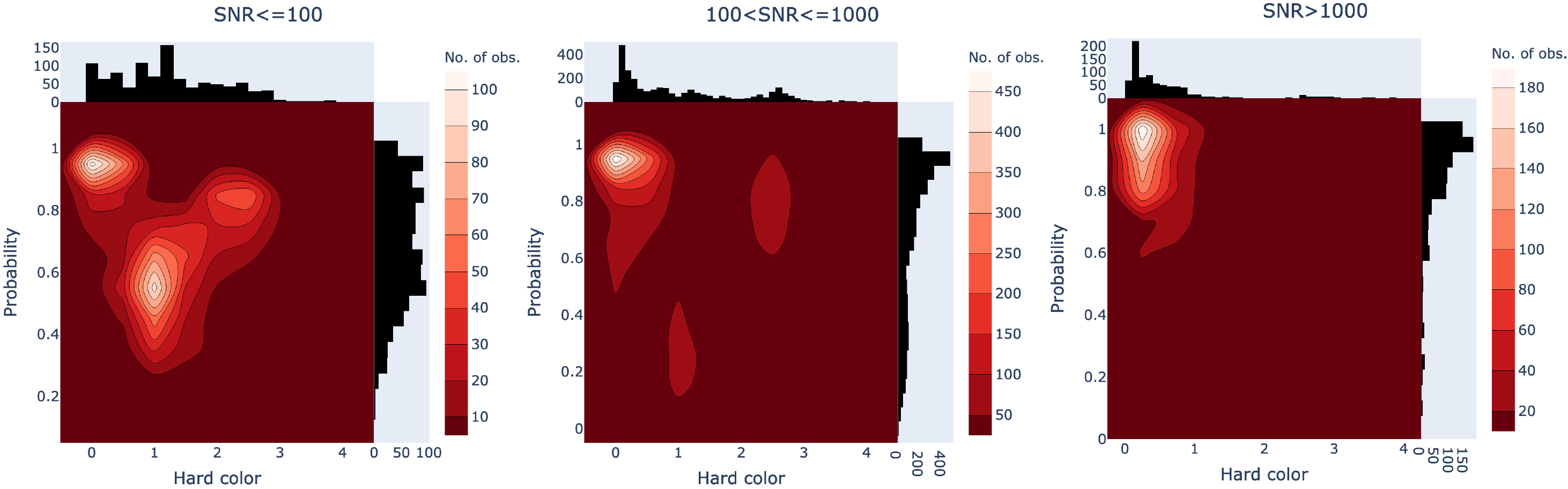}
}
\caption{Bi-variate density plots of Hard color and predicted probability of correct identification for NS and BH LMXB observations in different SNR ranges. The color bar shows the number of observations in each region. We have 1518, 3650 and 813 observations in the SNR less than 100, between 100 and 1000 and greater than 1000 ranges respectively for BH LMXBs. Similarly for NS LMXBs we have 1188, 6698 and 1018 observations in the SNR less than 100, between 100 and 1000 and greater than 1000 ranges respectively. The light-colored regions of the plots have the most number of observations. The individual uni-variate histogram plots for hard color and predicted probability are also shown on their respective axes. As can be observed, all the plots indicate that it's easier to classify observations with low hard color values and high SNR values.} 

\label{fig:hcvp}
\end{figure*}

In Fig.~\ref{fig:hcvp} we further investigate the correlation between hard color and predicted probability of correct identification of NS and BH LMXB observations for the different SNR ranges mentioned in section ~\ref{sec:snr_analysis}.  It follows the results shown in Fig.~\ref{fig:snrplt} and Fig.~\ref{fig:cci}. For the case of NS LMXBs we find that most of the observations in the low SNR range have predicted probability peaking around 0.5 and hard color value of 1.0. In the case of low SNR observations in BH LMXBs the predicted probabilities of most observations decreases as we increase the hard color value and then increases back again at hard color values \textgreater2. The same trend follows for BH LMXBs with SNR between 100 and 1000 although most observations this time have low hard color values and higher predicted probabilities. For BH LMXBs with SNR \textgreater1000 most observations have low hard color values and high predicted probabilities. In the case of the higher SNR ranges (\textgreater 100) for NS LMXBs most observations have predicted probability around 1.0 and hard color values \textless1.0. These plots again indicate that the algorithm can classify observations with low hard color values, i.e., HSS observations the best and the prediction accuracy increases with SNR. 

\subsection{Prediction for sample sources with unknown classification}
 
We use the final RF model trained on all 61 sources to predict the classification of a sample of 13 systems where the nature of the compact object is still unknown or under debate. These 13 sources were sampled with a total of 766 RXTE/PCA observations.
Our results and predictions  are summarized in table~\ref{tab:predtable}.

If \textgreater50\% of the observations of a source were predicted to belong to a particular class, that class was assigned to the source. Among the 13 sources, 5 sources (XTE J1901+014, XTE J1719--291, XTE J1727--476, IGR J17285--2922, XTE J1856+053) have very few observations (\textless10) that meet our criteria for good data (i.e. net count-rate \textgreater5 counts per second) and thus it is difficult for us to make any comments on the predicted classes for these sources. The remaining 8 sources (4U 1822--371, 4U 1957+11, IGR J17494--3030, SAX J1711.6--3808, SLX 1746--331, SWIFT J1842.5--1124, XTE J1637--498, XTE J1752--223) all have more than 30 observations each. Based on our criteria for classification mentioned earlier, 6 sources (4U 1822--371, 4U 1957+11, SLX 1746--331, SWIFT J1842.5--1124, XTE J1637--498, XTE J1752--223) were classified as BH LMXBs while 2 sources (IGR J17494--3030, SAX J1711.6--3808) were classified as NS LMXBs. 

Amongst the 8 sources with \textgreater30 observations, 5 sources have prediction percentage \textgreater60\%. Our model predicts that the source SAX J1711.6--3808 is a NS LMXB for ~94\% of its observations, however \cite{sanchez2006xmm} claim that SAX J1711.6--3808 might contain a black hole with a high spin parameter based on their fit of the X-ray spectra. For 88\% of its observations, the source SLX 1746--331 is predicted to have a BH, as speculated by \cite{white1996galactic} in their paper. Multiple works have argued that the compact object in 4U 1957+11 is a BH (\cite{nowak2011suzaku}, \cite{gomez2015case}) and our algorithm predicts the same for 72\% of its 121 observations. XTE J1752--223 is considered a BH LMXB candidate by \cite{shaposhnikov2010discovery} in their paper and our algorithm classified 67\% of its observations as a BH LMXB. The nature of the compact object in XTE J1637--498 is uncertain, but \cite{tetarenko2016watchdog} consider it as a BHC in their database. ~66\% of the observations of XTE J1637--498 is classified as BH LMXB by our algorithm. 

For the remaining 3 sources out of the aforementioned 8, the prediction percentage is \textless60\%, but still \textgreater50\%. For these sources we consider that the algorithm is confused about the nature of the compact object in the LMXBs. The source 4U 1822--371 is predicted to be a BH LMXB for 55\% of its observations but \cite{jonker2001discovery} detected pulsations from this source indicating that it most certainly is a NS LMXB.  \cite{armas2013xmm} have suggested that the source IGR J17494--3030 might be a NS LMXB and our trained model also predicts the same for 54\% of its 97 observations. SWIFT J1842.5--1124 was classified by \cite{zhao2016x} as a BH LMXB candidate and our trained model predicts it is a black hole for 51\% of its observations. Apart from that it is also important to note that all the 13 sources in our prediction sample have an average SNR \textless100 which is the region where the algorithm has the worst performance as shown in Fig.~\ref{fig:snrplt}.

\begin{table}
  \centering
  \resizebox{\columnwidth}{!}{
  \begin{tabular}{lcccc}
  	\toprule
Source name &  Total obs. &  Class &  Prediction (\%) & Avg. SNR\\
& & (Predicted) & &\\
\midrule
    \begin{filecontents*}{prediction_data_results.csv}
Source Name,Total Observations,Predicted Class,Prediction percentage,avg. snr
4U1822-371,97,BH,55.67,67.0
4U1957+11,121,BH,72.73,22.38
IGRJ17285-2922,5,BH,60.0,10.03
IGRJ17494-3030,97,NS,54.64,25.84
SAXJ1711.6-3808,34,NS,94.12,34.35
SLX1746-331,65,BH,87.69,26.82
SWIFTJ1842.5-1124,49,BH,51.02,25.71
XTEJ1637-498,76,BH,65.79,8.41
XTEJ1719-291,2,NS/BH,50.0,2.82
XTEJ1727-476,4,BH,100.0,6.3
XTEJ1752-223,210,BH,67.14,56.0
XTEJ1856+053,5,BH,100.0,10.75
XTEJ1901+014,1,BH,100.0,1.1
\end{filecontents*}
\csvreader[late after line=\\,head to column names]{prediction_data_results.csv}{1=\Source,2=\Total,3=\Class,4=\Percent,5=\SNR}{\Source & \Total & \Class & \Percent & \SNR}   
   	\bottomrule
  \end{tabular}}
  \caption{Classification results for sources in the prediction set. A class was assigned to a source if the majority of its observations were predicted to belong to that class. In cases where the ratio was 50-50 (XTE J1719-291) it is indicated that the source can belong to either class.}
  \label{tab:predtable}
\end{table}

\begin{figure}
    \includegraphics[width=\columnwidth,height=10cm]{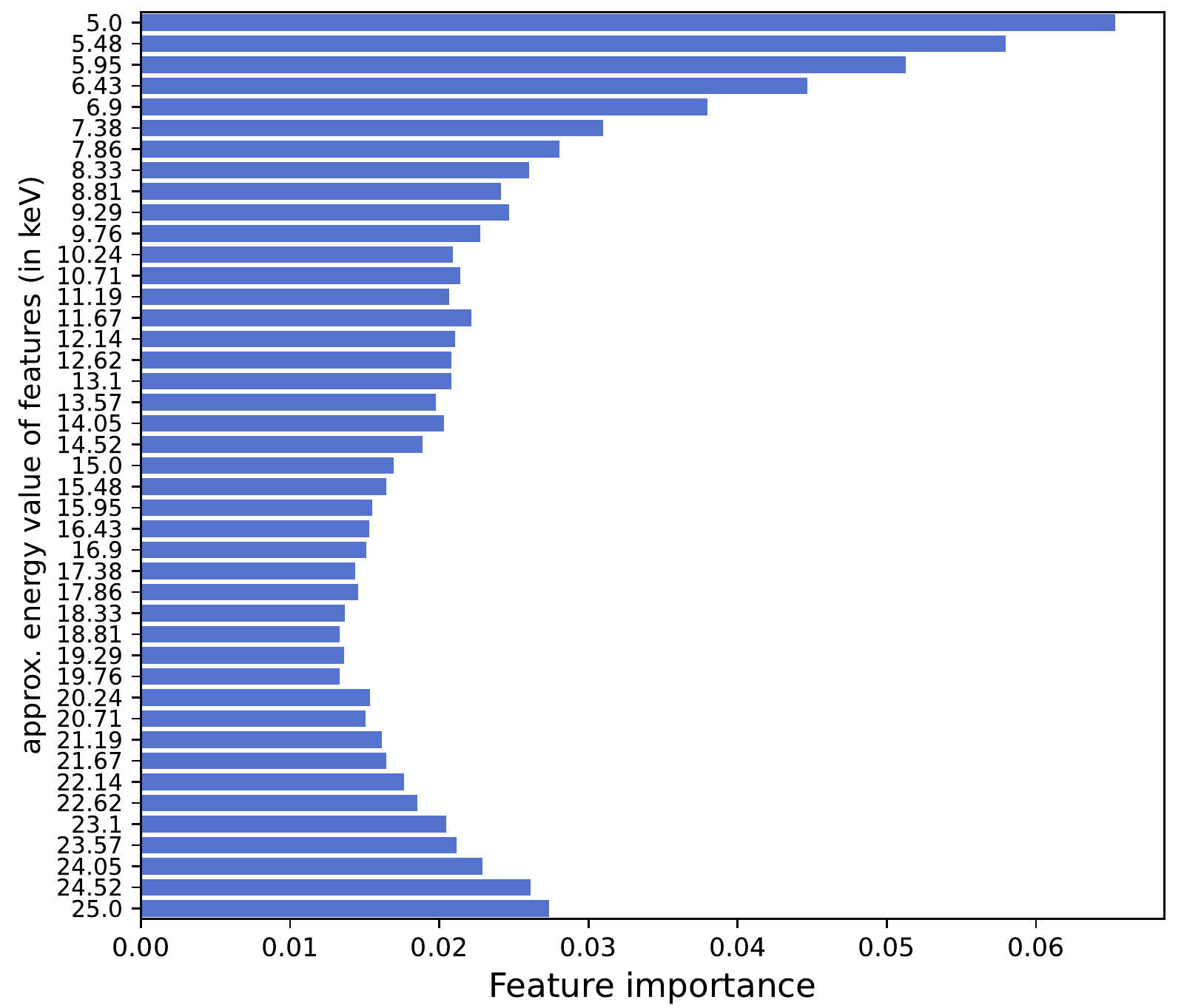}
    \caption{Feature importance plot for the input features. The y axis shows approximate energy values corresponding to each element in the input to the algorithm. The energy values increase from top to bottom. The x-axis has the feature importance of each element in the input vector to the algorithm. The sum of all importance values is equal to 1.}
    \label{fig:feature_imp}
\end{figure}

\section{Summary and Discussion}\label{sec:summary}

We used archival data from the PCA instrument aboard the RXTE mission (now decommissioned) to train a Random Forest algorithm that we subsequently use to classify a groups of Low Mass X-ray Binary (LMXB) systems into black-hole or neutron-star LMXB just by using their energy spectra as input. The data consist of 43 count rate values corresponding to the energy range of 5--25 keV for each observation of a source. The dataset consists of 14885 observations from 61 individual sources: 6216 observations from 28 BH systems and 8669 observations corresponding to 33 NS systems. We perform the training and testing using three different methods for a robust assessment of the performance of the RF algorithm for NS-BH classification. We obtain the outlier-resistant average model accuracy of $87\pm13\%$ at one sigma confidence level in classifying these systems. The final trained model is used to predict the classes of X-ray sources of unknown nature.

We also analyze the results of the classification by looking at the effect that signal-to-noise ratio and state transitions have on the predicted probabilities of correct identification. As expected it is observed that with better SNR the mean predicted probability of correct identification for observations increases. It is also observed that most of the observations (especially in the high SNR ranges) with a higher predicted probability have low hard color values and lie in the HSS.The higher predicted probability values of observations in the HSS can be attributed to their high SNR values. Another possible explanation to justify the better classification of observations in the HSS is the presence of a NS surface in the spectra of the HSS, which  would be absent in the HSS spectra of BH LMXBs.

To further investigate this, in Figure. ~\ref{fig:feature_imp} we plot the feature importance for the input spectra. The feature importance represents the relative importance that the ML algorithm gives to the given input data (in this case, flux at a given energy bin). Figure. ~\ref{fig:feature_imp} shows that both the lower-end and the higher-end of the spectra appear to be the most important parts of the energy spectra in order to differentiate between BH and NS. The least important part of the spectrum is around 18--19 keV, and around 12 keV there is a small bump suggesting that there might be weak features at this energy which also play an important role in the classification. 
The fact that Figure. ~\ref{fig:feature_imp} does not show a flat distribution is important, as it indicates that the algorithm is taking into account underlying differences in the energy spectra, which probably relate to subtle intrinsic physical differences between BH and NS (e.g. the presence of a surface and a boundary layer in the NS, potential differences in the size of the corona, contribution of the Jet to the X-rays, etc).

This, in turn, is pivotal to argue that if future works can use more interpretable class of algorithms  \cite[see, for e.g.,][and references therein]{villaescusa2020camels,udrescu2020ai} for this type of classification, then there is potential to use ML-techniques to learn more about the differences between BHs and NSs  from their spectral characteristics.

The main objective of this work was to probe whether machine learning techniques can be employed to determine the class of a newly observed LMXB source just by using the information contained in its energy spectra. Our results show that despite below average performance for a few sources, the random forest algorithm does a reasonably good job in classifying the NS-BH LMXBs overall. The most important aspect of this method is the speed of the classifications. Given an energy spectrum of a LMXB source, the algorithm is able to assign a class label to it in a fraction of a second. The algorithm also gives a probability of the predicted class for the spectrum that can be used as a confidence measure for the prediction. This algorithm has the potential of being used as a tool to very quickly flag the spectra of a newly identified source that can be helpful for scheduling follow-ups on particular objects of interest. It is also important to note that in most cases the net confidence of the predictions increases for a source as we add more observations.

One issue that we face currently in our work is that our classification model cannot be used directly to classify the energy spectra from other X-ray missions. The main reason for this is that most of the other currently active X-ray missions have instruments with effective areas that are different to RXTE's PCA. The first idea towards tackling this issue is to train a classification model for each instrument using their data. The problem that may arise while trying to do this is that there may not be enough data to train a machine learning algorithm for each instrument. That was one of the main reasons why we chose to work with data from RXTE even though it is now decommissioned. However the concept of transfer learning could be employed to train an algorithm for another instrument with limited data using our pre-trained classification model for RXTE data. More details on the idea behind transfer learning can be found in \citet{pan2009survey}.

Another alternative approach could be to use some sort of transformation to convert the data from a different instrument into the RXTE-PCA format. The transformed data can then be directly plugged into the pre-trained model. It is important, however, to realize that such a transformation is only possible for data obtained from instruments that have overlapping operational energy range (i.e. at least 5-25 keV). This rules out data obtained from instruments that operate specifically at lower energy ranges (e.g. the SWIFT's X-Ray Telescope) as we do not use data below 5 keV to avoid any effect of interstellar absorption.\\ 

 Adding more information as input to the algorithm can also be explored as a means of improving the current level of accuracy reached for all the sources in our dataset. One way of doing that would be to combine the energy spectra with the power spectra of all observations for each source. There are many more potential directions that can be explored in the future for solving the problem of Low-Mass X-ray Binary spectral classification. We believe that our experiment can serve as a starting point for the application of machine learning methods to solve this and other problems in the domain of X-ray astronomy.

\section*{Acknowledgements}
We acknowledge the financial support from the Royal Society, from a Raja Ramanna Fellowship awarded by Department of Atomic Energy (DAE), India (10/1(16)/2016/RRF-R\&D-II/630) and from an INSPIRE Faculty fellowship research grant (IFA16-PH176) awarded by Department of Science and Technology (DST), India. We thank Phil A. Charles for helping us in determining the class (BH/NS) of a number of LMXB using available data. We thank Abhirup Datta, Jamie Court,  Kaustubh Vaghmare, Adam Hill, Gulab Chand Dewangan and Ranjeev Misra for insightful discussion. KA acknowledges support from a UGC-UKIERI Phase 3 Thematic Partnership
(UGC-UKIERI-2017-18-006; PI: P. Gandhi).

\section*{Data availability}

The data underlying this article are publicly available in the High Energy Astrophysics Science Archive Research Center (HEASARC) at \url{https://heasarc.gsfc.nasa.gov/db-perl/W3Browse/w3browse.pl}




\bibliographystyle{mnras}
\nocite{*}
\bibliography{paper} 


\appendix

\onecolumn
\section{Complete tables of source wise accuracies for Method 2 and Method 3}

\begin{filecontents*}{Method-2-Source-wise-Accuracy.csv}
Name,Class,Total Observations,Correctly Classified,Misclassified,Accuracy
XTEJ1118+480,BH,80,11,69,13.75
XTEJ1748-288,BH,91,31,60,34.07
XTEJ1652-453,BH,55,35,20,63.64
XTEJ1759-220,NS,45,29,16,64.44
SWIFTJ1756.9-2508,NS,47,33,14,70.21
MAXIJ1836-194,BH,74,52,22,70.27
SWIFTJ1713.4-4219,BH,31,24,7,77.42
SLX1735-269,NS,83,65,18,78.31
GRS1737-31,BH,14,11,3,78.57
NGC6440,NS,87,74,13,85.06
4U1543-47,BH,67,57,10,85.07
SAXJ1819.3-2525,BH,9,8,1,88.89
4U1746-371,NS,61,55,6,90.16
SWIFTJ1357.2-0933,BH,23,21,2,91.3
SAXJ1810.8-2609,NS,36,33,3,91.67
SAXJ1806.5-2215,NS,50,46,4,92.0
IGRJ17497-2821,NS,13,12,1,92.31
KS1731-260,NS,75,72,3,96.0
MXB1658-298,NS,73,71,2,97.26
1A1744-361,NS,49,48,1,97.96
XTEJ1755-324,BH,10,10,0,100.0
XTEJ2012+381,BH,26,26,0,100.0
4U1254-690,NS,100,100,0,100.0
GS1354-64,BH,11,11,0,100.0
GRS1739-278,BH,11,11,0,100.0
V4641SGR,BH,7,7,0,100.0
XTEJ1818-245,BH,56,56,0,100.0
\end{filecontents*}
\csvreader[
  longtable=lccccc,
  table head=\caption{Source wise performance of algorithm in method 2 (source-wise train-test split).}\label{tab:acctable1}\\
    \toprule Source Names &  Class &  Total test obs. &  Correctly Classified &  Misclassified &  Accuracy percentage \\ \midrule\endhead
    \bottomrule\endfoot,
  late after line=\\,
]{Method-2-Source-wise-Accuracy.csv}{1=\Source,2=\Class,3=\Total,4=\Correct,5=\Miss,6=\Acc,}{\Source & \Class & \Total & \Correct & \Miss & \Acc}

\begin{filecontents*}{Method-3-Source-wise-Accuracy.csv}
Name,Class,Total Observations,Correctly Classified,Misclassified,Accuracy
XTEJ1118+480,BH,80,6,74,7.5
IGRJ00291+5934,NS,180,14,166,7.78
1A1246-588,NS,166,36,130,21.69
XTEJ1748-288,BH,91,27,64,29.67
IGRJ17379-3747,BH,784,415,369,52.93
XTEJ1812-182,NS,233,129,104,55.36
XTEJ1908+094,BH,213,121,92,56.81
4U1728-34,NS,423,264,159,62.41
XTEJ1652-453,BH,55,35,20,63.64
H1743-32,BH,558,361,197,64.7
XTEJ1759-220,NS,45,30,15,66.67
SAXJ1808.4-3658,NS,295,206,89,69.83
SWIFTJ1756.9-2508,NS,47,33,14,70.21
MAXIJ1836-194,BH,74,52,22,70.27
SWIFTJ1539.2-6227,BH,145,103,42,71.03
SWIFTJ1713.4-4219,BH,31,24,7,77.42
GRS1737-31,BH,14,11,3,78.57
GRS1747-312,NS,215,170,45,79.07
TERZAN5,NS,125,99,26,79.2
LMCX-2,NS,141,112,29,79.43
4U1630-47,BH,1102,877,225,79.58
XTEJ1720-318,BH,101,82,19,81.19
MAXIJ10556-332,NS,262,217,45,82.82
4U1608-52,NS,1041,876,165,84.15
4U1543-47,BH,67,58,9,86.57
SLX1735-269,NS,83,72,11,86.75
CYGX-2,NS,583,514,69,88.16
NGC6440,NS,87,77,10,88.51
SAXJ1819.3-2525,BH,9,8,1,88.89
4U1746-371,NS,61,55,6,90.16
GX339-4,BH,1163,1055,108,90.71
XTEJ1650-500,BH,121,110,11,90.91
XTEJ1550-564,BH,368,335,33,91.03
SWIFTJ1357.2-0933,BH,23,21,2,91.3
AQLX1,NS,555,507,48,91.35
MAXIJ1543-564,BH,268,245,23,91.42
SAXJ1806.5-2215,NS,50,46,4,92.0
XTEJ1859+226,BH,125,115,10,92.0
IGRJ17497-2821,NS,13,12,1,92.31
4U0614+091,NS,498,464,34,93.17
SAXJ1810.8-2609,NS,36,34,2,94.44
XTEJ1817-330,BH,157,149,8,94.9
HETEJ1900.1-2455,NS,351,336,15,95.73
GROJ1655-40,BH,546,524,22,95.97
4U1705-44,NS,512,492,20,96.09
MXB1658-298,NS,73,71,2,97.26
1A1744-361,NS,49,48,1,97.96
4U1724-307,NS,127,125,2,98.43
SAXJ1750.8-2900,NS,131,129,2,98.47
KS1731-260,NS,75,74,1,98.67
4U1636-53,NS,1563,1555,8,99.49
4U1254-690,NS,100,100,0,100.0
V4641SGR,BH,7,7,0,100.0
XTEJ1755-324,BH,10,10,0,100.0
XTEJ2012+381,BH,26,26,0,100.0
GS1354-64,BH,11,11,0,100.0
XTEJ1818-245,BH,56,56,0,100.0
SERX-1,NS,102,102,0,100.0
4U1702-429,NS,225,225,0,100.0
GRS1739-278,BH,11,11,0,100.0
4U1735-44,NS,222,222,0,100.0
\end{filecontents*}
\csvreader[
  longtable=lccccc,
  table head=\caption{Source wise performance of algorithm in method 3 (leave-one source out).
}\label{tab:acctable2} \\
    \toprule Source Names &  Class &  Total test obs. &  Correctly Classified &  Misclassified &  Accuracy percentage \\ \midrule\endhead
    \bottomrule\endfoot,
  late after line=\\,
]{Method-3-Source-wise-Accuracy.csv}{1=\Source,2=\Class,3=\Total,4=\Correct,5=\Miss,6=\Acc,}{\Source & \Class & \Total & \Correct & \Miss & \Acc}



\label{lastpage}
\end{document}